\documentclass[aps,prd,10pt,twocolumn,nofootinbib,nobibnotes,longbibliography,superscriptaddress]{revtex4-2}

\usepackage[utf8]{inputenc} % Allows direct input of ñ, á, é, etc.
\usepackage[T1]{fontenc}  % Correct accent rendering in output
\usepackage{lmodern}
\usepackage{graphicx}
\usepackage{amsmath,amssymb}
\usepackage{hyperref}
\usepackage{color}
\usepackage{physics}
\usepackage{enumerate}
\usepackage{overpic}
\usepackage[dvipsnames]{xcolor}

\hypersetup{
    colorlinks=true,
    linkcolor=blue,
    citecolor=blue,
    urlcolor=blue,
    pdftitle={Your Review Title},
    pdfauthor={Your Name}
}

\usepackage[normalem]{ulem}

\graphicspath{{images/}}  

\begin{document}

\title{Gravity from equilibrium thermodynamics of stretched light cones}
  
\author{Ana Alonso-Serrano}
\email{ana.alonso.serrano@aei.mpg.de}
\affiliation{\mbox{Institut für Physik, Humboldt-Universität zu Berlin, Zum Großen Windkanal 6, 12489 Berlin, Germany}}
\affiliation{Max-Planck-Institut f\"ur Gravitationsphysik (Albert-Einstein-Institut), Am M\"{u}hlenberg 1, 14476 Potsdam, Germany}

\author{Luis J. Garay}
\email{luisj.garay@ucm.es}
\affiliation{\mbox{Departamento de F\'{i}sica Te\'{o}rica and IPARCOS, Universidad Complutense de Madrid, 28040 Madrid, Spain}}

\author{Marek Liška}
\email{liskama4@stp.dias.ie}
\affiliation{School of Theoretical Physics, Dublin Institute for Advanced Studies, 10 Burlington Road, Dublin, Ireland}

\author{Celia López Pineros}
\affiliation{\mbox{Departamento de F\'{i}sica Te\'{o}rica and IPARCOS, Universidad Complutense de Madrid, 28040 Madrid, Spain}}

\begin{abstract}
This work digs into the connection between gravity and thermodynamics of stretched light cones (SLC). They are associated with uniformly accelerating observers, who endow the SLC with a physical notion of temperature via the Unruh effect. We compute the expansion, shear, and vorticity of the SLC to fully study its dynamics and account for the possibility of previously predicted non-equilibrium entropy production. For consistency, we prove the equivalence of the two different geometrical methods available for studying the SLCs' properties. Then, we apply the energy balance and use Clausius' relation to relate the geometrical properties of the SLC with energy fluxes crossing its surface, showing that it encodes the equations governing the gravitational dynamics. We show here how this analysis can be fully carried in terms of equilibrium thermodynamics due to the vanishing of shear, and how one can identify a work term related to the acceleration of the observer.

\end{abstract}

\maketitle

\tableofcontents
\section{Introduction}\label{Sec:Introduction}

The profound connection between gravity and thermodynamics in physics has become more and more significant over the years. From the laws of black hole mechanics \cite{Bekenstein:1973,Bardeen:1973,Hawking} (that surprisingly resemble the three laws of thermodynamics), to the Unruh effect \cite{Unruh} (where an accelerating observer detects a thermal spectrum in what inertial observers perceive as vacuum), several studies point toward a strong thermodynamic nature of the spacetime itself. These initial ideas led to posing a very strong and interesting question: can Einstein's equations, that explain the dynamics of spacetime, be also obtained as an equation of state, resulting from working with the thermodynamical formalism? If so, what would be the true nature of spacetime's entropy, and how could we relate it to the energy and matter in it? Solving these questions would lead not only to a better understanding of spacetime itself, but also of its thermodynamic nature as such. 

One of the most influential proposals in this formalism, is that of Jacobson~\cite{Jacobson_1995}, where Einstein's equations were derived as local constitutive equations for an equilibrium spacetime. This study was specifically done for local Rindler causal horizons (those seen by accelerating observers, that delimit the region of spacetime  causally accessible to them), under the assumption that their entropy is proportional to the area of the horizons. Since then, many researchers have explored the connection between thermodynamics and spacetime.

Following Jacobson's results, further research has been done extending his results both conceptually and mathematically. Notably, it has been shown that thermodynamic equilibrium can recover not only Einstein equations but even the equations of motion of a wide class of modified theories of gravity~\cite{Eling:2006,Padmanabhan:2010,Jacobson:2012,Mohd_2016,Parikh_2018,Alonso:2024}. Thermodynamic perspective on gravitational dynamics also continues to work beyond the context of classical gravity, as semiclassical gravitational equations follow from equilibrium conditions imposed on entanglement entropy~\cite{Faulkner:2014,Jacobson_2016,Faulkner:2017,Svesko:2019,Kumar:2023}. On the mathematical side, the expression for the matter entropy flux across a causal horizon has been put on a more solid ground~\cite{Baccetti:2013ica}, and deformations of local causal horizons in curved spacetimes have been explored in greater detail~\cite{Jacobson:2017,Wang:2019}.

In particular, two developments in thermodynamics of spacetime serve as an inspiration for this work. First, Chirco and Liberati \cite{Chirco_2010} have examined the implications of non-equilibrium thermodynamics in the context of gravity. They have done so by allowing for entropy production encoded in non-vanishing shear contributions in Raychaudhuri's equation \cite{Raychaudhuri:1955,Raych}. They argue that Einstein's equations emerge not from strict equilibrium conditions, but from a more general entropy balance law inside the non-equilibrium thermodynamics framework, that accounts for dissipation processes. 

More recently, Parikh and Svesko \cite{Parikh_2018} introduced a new geometric setting for the study: the so called stretched light cone (SLC), as the scene for deriving gravitational field equations. Considering a congruence formed by uniformly accelerating observers that defines a timelike hypersurface near any spacetime point, they use the SLC in the same way that Chirco and Liberati used Rindler horizons. In their approach, carried out in the limit of small time and length scales, they derive Einstein’s equations by relating the deviation of the congruence's generators  and the heat fluxes across the SLC surface. An essential aspect in their method is, again, the use of non-equilibrium thermodynamics. 

In this work, we aim to connect both approaches, and physically analyze the results. In order to do so, we study non-geodesic, accelerated congruences generated by uniformly accelerating observers, associated with a SLC. We employ their corresponding Raychaudhuri's equation, and analyze its content to extract the physical properties of the system, including those born from its non-geodesic nature. Specifically, we consider the additional acceleration-dependent terms that lead to extra geometric contributions, especially, and most notably, the divergence of the acceleration vector (that was not present in previous derivations focused on null surfaces \cite{Jacobson_1995,Chirco_2010}). This introduces a natural pathway to interpret the thermodynamics of SLCs solely in terms of equilibrium processes, in contrast to the previous works  \cite{Chirco_2010,Parikh_2018}. This equilibrium description is possible, since, as we will see, the dissipation term related to the shear vanishes for our system, and the previously reported irreversible entropy production in SLCs can be interpreted as a work term.

In addition and in contrast to previous studies, which restrict themselves to either direct geometric computation \cite{Parikh_2018} or Raychaudhuri’s evolution equation \cite{Jacobson_1995,Chirco_2010} separately, in this work, we perform a comparative analysis of both approaches. We also demonstrate that, when applied to SLCs with the appropriate constraints (e.g., vanishing initial vorticity and expansion), both methods yield consistent physical results for the expansion. A crucial aspect is the entropy-area relation \cite{SA,SAA}. It not only makes explicit the thermodynamic interpretation of the spacetime, but shows the relation between geometry, acceleration, and entropy production.

To finish our study, by integrating the identity obtained by combining both Clausius', and the entropy-area relation in a generalized equilibrium setup now including the extra acceleration-dependent term (which we justify in the corresponding section), we derive Einstein’s equations from this new thermodynamic setting. The interpretation of the acceleration-dependent contributions is highly non-trivial. Previous works identified similar terms as a non-equilibrium, observer-dependent, entropy flux in dynamical spacetimes. However, we argue that they instead encode external work required to maintain the Rindler observers' acceleration. Therefore, this study not only generalizes the derivation of gravitational dynamics from thermodynamics, but it also unlocks new questions that will help us rethink the role of accelerating observers and the physical meaning of entropy in curved spacetimes.

In summary, this work aims to unify and extend previous works like that of Jacobson \cite{Jacobson_1995}, Chirco and Liberati \cite{Chirco_2010}, and Parikh and Svesko \cite{Parikh_2018}, within a joined framework of equilibrium and non-geodesic gravitational thermodynamics. Through the explicit use of non-geodesic congruences and comparison of alternative methods, it offers a refined understanding of how the gravitational field equations are encoded in the observer-dependent thermodynamical setting.

The paper is organized as follows. In section \ref{section: raych}, we introduce Raychaudhuri's equation for non-geodesic congruences. This allows us to study their evolution. In section \ref{section: geom}, we describe the geometrical setting we work with. We build the SLC in subsection \ref{sub:SLC}, and in subsection \ref{sec:omsigex} we compute the expansion, shear, and vorticity for it, via both approaches mentioned (geometric computation and Raychaudhuri’s evolution equation). We then compare both results. In subsection~\ref{sub:area}, we use the expansion to compute the area of the SLC. In section \ref{section: energy}, we explore the thermodynamical nature of heat and work in this relativistic system. Using these concepts, section \ref{section: therm} realizes the thermodynamical analysis of the system, which allows us to obtain the sought-for Einstein's equations. Finally, in section \ref{section: conclusions} we draw our conclusions, and open the doors to potential future studies by posing new questions.

\section{Raychaudhuri's equation for non-geodesic congruences}\label{section: raych}

Raychaudhuri's equation provides a fundamental description of how congruences (characterized by a tangent vector field) evolve in a given spacetime.  Therefore, in this section we are going to briefly review how this equation behaves for non-geodesic congruences (the ones swept out by accelerating observers). As we show in the following, acceleration-related term appears in Raychaudhuri's equation and will become important when studying the gravitational dynamical equations via thermodynamic methods. Raychaudhuri's equation also introduces the description of the congruences in terms of expansion, shear, and vorticity, relevant for the thermodynamical analysis. As we will see, the expansion encodes entropy changes, and vanishing vorticity implies that the congruence is orthogonal to hypersurfaces.

Let us consider a congruence characterized by a timelike normalized tangent vector, $u^\mu$ (such that \mbox{$u_\mu u^\mu=-1$}). Since we are working with a non-geodesic congruence, in general, \mbox{$u^\nu\nabla_\nu u^\mu=a^\mu\neq0$}. Raychaudhuri's equation concerns the evolution of the deviation tensor $\nabla_\nu u_\mu$, which encodes the changes of the velocity field. We can decompose it as
\begin{align}
\begin{split}
 \nabla_\nu u_\mu=\omega_{\mu\nu}+\sigma_{\mu\nu}+\frac{1}{3}\theta h_{\mu\nu}-u_{\nu}a_{\mu},\label{equation: newraych}
\end{split}
\end{align}
where $h_{\mu\nu}=g_{\mu\nu}+u_{\mu}u_{\nu}$ is the projector to the plane orthogonal to $u^{\mu}$. In this equation, the vorticity $\omega_{\mu\nu}$ is an antisymmetric tensor that measures the rotation or twirling of the velocity field
\begin{align}
\begin{split}
\omega_{\mu\nu}=h_{[\mu}^{\rho}h_{\nu]}^{\lambda}\nabla_{\lambda}u_{\rho}
=\nabla_{[\nu}u_{\mu]}+u_{[\nu}a_{\mu]}\label{equation: vort};
\end{split}
\end{align}
the shear $\sigma_{\mu\nu}$ is a symmetric traceless tensor that measures how the velocity is distorted or changes the congruence in shape
\begin{equation}
\sigma_{\mu\nu}=\theta_{\mu\nu}-\frac{1}{3}\theta h_{\mu\nu}=\nabla_{(\nu}u_{\mu)}+u_{(\nu}a_{\mu)}-\frac{1}{3}\theta h_{\mu\nu}; \label{equation: shear}
\end{equation}
and the expansion $\theta$ is a scalar that accounts for the change of the congruence's cross-section area
\begin{align}
\begin{split}
\theta=\nabla_\mu u^\mu.\label{equation: exp}
\end{split}
\end{align} 
Note that, when $a^\mu=0$ (i.e. in the geodesic case), we recover the standard definitions for geodesic congruences. 

The identity
\begin{align}
\begin{split}
 u^\rho\nabla_\rho\nabla_\nu u_\mu=\nabla_\nu a_\mu-\nabla_\nu u^\rho\nabla_\rho u_\mu+R_{\rho\nu\mu\sigma}u^\rho u^\sigma \label{equation: evdevvect}
\end{split}
\end{align}
then allows us to obtain the evolution equation for the expansion scalar as the trace of the evolution of the deviation vector plus the evolution of the trace of the acceleration terms
\begin{align}
 \dot{\theta}&= u^\nu\nabla_\nu\nabla_\mu u^\mu 
 \nonumber\\
 &=\nabla_\mu a^\mu+\omega^2-\sigma^2-\frac{1}{3}\theta^2-R_{\rho\sigma}u^\rho u^\sigma.\label{equation: evexp}
\end{align}
This result is the Raychaudhuri's equation. Compared to the geodesic Raychaudhuri's equation \cite{Raych}, it includes an extra divergence of the acceleration. The presence of this term results in a new quantity to take into account when applying the thermodynamical analysis in contrast with previous works (e.g. \cite{Chirco_2010}).

\section{Geometrical setting}\label{section: geom}

As discussed in the introduction, one of the essential pieces in this study is the proper definition of the geometrical concepts that will be used throughout. We have already introduced Raychaudhuri's equation for non-geodesic congruences in the previous section. We now focus on the construction of the SLC and the definition of the expansion $\theta$, the shear $\sigma_{\mu\nu}$, and the vorticity $\omega_{\mu\nu}$ for SLCs. 

\subsection{Construction of stretched light cones}\label{sub:SLC}

We first construct a SLC in flat spacetime as a surface generated by wordlines of radial observers with constant acceleration. To define it, consider a timelike congruence generated by a unit, timelike, spherical vector field $u^{\mu}$ such that
\begin{equation}
u^\mu_{f}=\left(\frac{r}{\sqrt{r^2-t^2}},\frac{t}{r\sqrt{r^2-t^2}}x,\frac{t}{r\sqrt{r^2-t^2}}y,\frac{t}{r\sqrt{r^2-t^2}}z \right), \label{uflat}
\end{equation}
where we have defined $r=\sqrt{x_ix^i}$ as the radial distance from $p$, and the index $i=x,y,z$ just corresponds to the spatial Cartesian coordinates.

\begin{figure}[tbp]
\centering
\includegraphics[width=0.75\columnwidth,origin=c,trim={4.5cm 15.5cm 4.95cm 1.9cm},clip]{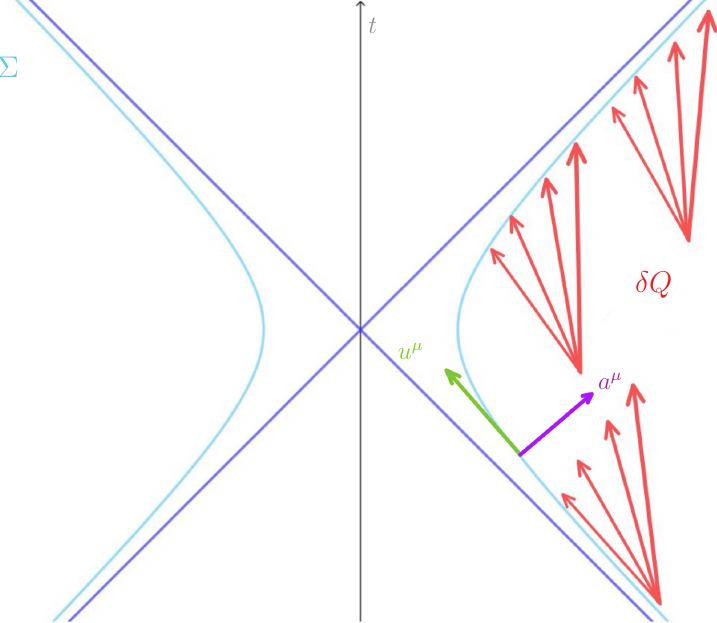}
\caption{\small Graphical depiction of a SLC $\Sigma$, the velocity field $u^\mu$ defined on it, its acceleration $a^\mu$, and the heat flux crossing its surface $\delta Q$. The darker blue lines denote the null surface which the SLC approaches in the limit of infinite acceleration, i.e. the light cone.}
\label{fig:Fluxx}
\end{figure}

We can define a 3-dimensional timelike hyperboloid $\Sigma$, as depicted in FIG.~\ref{fig:Fluxx}, defined by the set of wordlines tangent to $u^{\mu}$ that satisfy
\begin{equation}
r^2-t^2=\alpha^2 ,\label{slcmink}
\end{equation}
with $\alpha$ a constant with dimensions of length. This is precisely the equation that describes the surface of the SLC. For $\alpha=0$, we recover the light cone. Now the surfaces satisfying equation \eqref{slcmink} at constant $t$ are nothing but 2-dimensional spheres with an area given by
\begin{equation}
A(t)=4\pi(\alpha^2+t^2). \label{areaSLC}
\end{equation}
We can easily verify that the velocity field tangent to a SLC has uniform acceleration. Indeed, the proper acceleration
\begin{equation}
a^\mu_f=u^\nu_f\nabla_\nu u^\mu_f, \label{acceldef}
\end{equation}
has a modulus on $\Sigma$ equal to $a_f=1/\alpha$.

We now extend the definition of the SLC to a general curved spacetime. To establish the notion of approximate spherical boosts, we specify the metric via its expansion in the Riemann normal coordinates centered at any point $p$ of a given spacetime, such that in a normal neighborhood of $p$, the metric is locally equivalent to Minkowski's, as follows \cite{Riemann}
\begin{equation}
\label{RNC}
g_{\mu\nu}(x)=\eta_{\mu\nu}-\frac{1}{3}R_{\mu\rho\nu\sigma}(p)x^\rho x^\sigma+\mathcal{O}(x^3),
\end{equation}
where $x^\mu$ are the normal coordinates and $\mathcal{O}(x^3)$ stands for higher order terms in $x$. For simplicity in notation and calculations, we define the quantity
\begin{equation}
\label{p tensor}
p_{\mu\nu}=\frac{1}{3}R_{\mu\rho\nu\sigma}(p)x^\rho x^\sigma,
\end{equation}
which we will use throughout. It is also useful to compute the Christoffel symbols that read
\begin{equation}
\Gamma^\gamma_{\mu\nu}=-\frac{1}{3}\left[R_{\ \mu\nu\lambda}^\gamma(p)+R_{\ \nu\mu\lambda}^\gamma(p)\right]x^\lambda+\mathcal{O}(x^2)  .\label{christoffels}
\end{equation}

To construct a SLC, we now want the hyperboloid~$\Sigma$ to be defined in such a way that it possesses a constant and uniform temperature locally, so we can proceed with the upcoming thermodynamical study of it. This condition is satisfied if $\Sigma$ is formed by worldlines of uniformly accelerating observers. However, due to the spacetime's curvature, it does not really correspond to worldlines with constant acceleration. To account for this effect, we select a set of trajectories that at $t=0$ have an instantaneous proper acceleration given by $a=1/\alpha$. Following said trajectories along time, the acceleration varies but, if we choose a time scale $\epsilon\ll\alpha$, and $\alpha$ much smaller than the local curvature length scale (the inverse square root of the largest eigenvalue of the Riemann tensor), we can regard the acceleration as approximately constant. Every calculation will then be made in the time scale given by  $t\leq\epsilon\ll\alpha$. 

We again define the SLC as the timelike surface $\Sigma$ at $r^2-t^2=\alpha^2$ with $u^{\mu}$ being tangent to it. Now, in the same way that $\Sigma$ is no longer a strict hyperboloid as in the Minkowski case, its cross-sections are no longer exactly 2-spheres. However, this variation is very small during the time scale $\epsilon$, and does not affect the temperature.

Finally, motivated by the Unruh effect \cite{Unruh}, we can assign a temperature to $\Sigma$ given by
\begin{align}
T_{\text{U}}=\frac{\hbar a}{2\pi}=\frac{\hbar}{2\pi\alpha}. \label{Unruh}
\end{align}
In flat spacetime, this expression is exact. In a curved spacetime, $T_{\text{U}}$ develops curvature-dependent corrections, which have no effect on our study (including all the time-dependent corrections to $T_{\text{U}}$). This temperature, was originally derived in flat spacetime for eternally uniformly accelerating observers \cite{Unruh}. However, for sufficiently large acceleration $a$, the result also applies for finite lifetime observers in generic curved spacetimes \cite{Barbado_2012}. In our case, choosing $t\leq\epsilon\ll\alpha$ means that we can consider our observer and system to be thermalized, and, therefore, to share the same temperature $T_{\text{U}}$.

\subsection{\texorpdfstring{$\theta$}{TEXT}, \texorpdfstring{$\sigma_{\mu\nu}$}{TEXT}, and \texorpdfstring{$\omega_{\mu\nu}$}{TEXT} for SLCs}\label{sec:omsigex}

We compute here the expansion, shear, and vorticity for the SLCs, following the definitions in the previous section in two different approaches. The first approach is to compute $\theta$, $\sigma_{\mu\nu}$, and $\omega_{\mu\nu}$ directly from the velocity field; the second approach computes them, given some appropriate initial conditions, via the Raychaudhuri's equation. Let us emphasize that we restrict to first order in curvature and small times, i.e. $t\le\epsilon\ll\alpha$. The motivation behind proceeding in both ways is that in previous works, this kind of studies have been carried out either in one way (using the velocity field \cite{Parikh_2018}) or another (with the Raychaudhuri approach \cite{Jacobson_1995}, \cite{Chirco_2010}) respectively. In this section, we aim to compare both approaches and show that they are, in fact, equivalent. 

To proceed, we first need to understand the curved spacetime construction of velocity $u^\mu$. Using the Riemann normal coordinate system, we can easily introduce the flat spacetime velocity $u^\mu_f$ given by equation~\eqref{uflat} in any regular point $p$ of a generic curved spacetime\footnote{The introduction of $u^\mu_f$ in a curved spacetime does not depend on using the Riemann normal coordinates in any way, they just make its construction particularly convenient.}. However, due to curvature-dependent corrections to the metric (captured by equation~\eqref{RNC}) $u^\mu_f$ only satisfy the normalization condition $g_{\mu\nu}u^\mu_fu^\nu_f=-1$, directly in $p$, and not in its vicinity. The total velocity $u^{\mu}$ must then also include curvature-dependent corrections. Therefore, we define it as follows
\begin{align}
u^\mu=u^\mu_f+u^\mu_c,
\label{vel}
\end{align}
where $u^\mu_f$  is the flat contribution, and $u^\mu_c$ is an unknown curved contribution that vanishes in flat spacetime. We need to define $u^\mu_c$ in a way that keeps the complete velocity field in equation~\eqref{vel}, normalized to $g_{\mu\nu}u^{\mu}u^{\nu}=-1$.  
Since we are only interested in studying the congruence generated by $u^{\mu}$ for small values of $t$, we can conveniently expand $u^{\mu}$ in powers of $t/\alpha\le\epsilon/\alpha\ll1$, keeping up to second order terms, i.e.: 
\begin{align}
u^\mu=u^\mu_0+u^\mu_1\frac{t}{\alpha}+u^\mu_2\frac{t^2}{2\alpha^2}+\mathcal{O}\left(\frac{t^3}{\alpha^3}\right), \label{veldef}
\end{align}
where we separate each component according to~\eqref{vel} as $u^\mu_n=u_{nc}^\mu+u_{nf}^\mu$ with $n=0,1,2$.
For the acceleration we keep the expansion up to the first power in time (as it is defined with derivatives of the velocity, and $\mathcal{O}(t^2/\alpha^2)$ terms depend on $\mathcal{O}(t^3/\alpha^3)$ contributions from the velocity) as
\begin{align}
    a^\mu=a^\mu_0+a^\mu_1\frac{t}{\alpha}+\mathcal{O}\left(\frac{t^2}{\alpha^2}\right), \label{acceltime}
\end{align}
where, once again, we separate each component according to \eqref{vel} as $a^\mu_n=a_{nc}^\mu+a_{nf}^\mu$ with \mbox{$n=0,1$}. Applying the definition (\ref{acceldef}), the flat contribution obeys
\begin{equation}
a^\mu_{0f}=\left(0,\frac{x}{r^2},\frac{y}{r^2},\frac{z}{r^2}\right),\qquad a^\mu_{1f}=\left(\frac{\alpha}{r^2},0,0,0\right). \label{accelflat}
\end{equation}

Next, we discuss the expansion of the metric and the Christoffel symbols in powers of $t$. For the metric, we can expand $p_{\mu\nu}$ defined in equation~\eqref{p tensor} as follows
\begin{equation}
p_{\mu\nu}=p_{0\mu\nu}+p_{1\mu\nu}\frac{t}{\alpha}+p_{2\mu\nu}\frac{t^2}{2\alpha^2}+O\left(\frac{t^3}{\alpha^3}\right),
\end{equation}
with
\begin{align}
p_{0\mu\nu}&=\frac{1}{3}R_{\mu i\nu j}\left(p\right)x^ix^j,\nonumber\\  
p_{1\mu\nu}&=\frac{\alpha}{3}\left[R_{\mu i\nu t}\left(p\right)+R_{\mu t\nu i}\left(p\right)\right]x^i,\\ 
p_{2\mu\nu}&=\frac{2\alpha^2}{3}R_{\mu t\nu t}\left(p\right)\nonumber. 
\end{align} 
For the Christoffel symbols, we can proceed analogously taking their definition \eqref{christoffels} so that
\begin{equation}
\Gamma^\gamma_{\mu\nu}=\Gamma^\gamma_{0\mu\nu}+\Gamma^\gamma_{1\mu\nu}\frac{t}{\alpha}+O\left(\frac{t^2}{\alpha^2}\right),
\end{equation}
with
\begin{align}
\Gamma^\gamma_{0\mu\nu}&=-\frac{1}{3}\left[R_{\ \mu\nu i}^\gamma(p)+R_{\ \nu\mu i}^\gamma(p)\right]x^i,\nonumber\\ 
\Gamma^\gamma_{1\mu\nu}&=-\frac{\alpha}{3}\left[R_{\ \mu\nu t}^\gamma(p)+R_{\ \nu\mu t}^\gamma(p)\right].
\end{align}

Finally, we expand the expansion $\theta$, the shear $\sigma_{\mu\nu}$, and the vorticity $\omega_{\mu\nu}$ up to the first power in time because the next order is, again, affected by $\mathcal{O}(t^3/\alpha^3)$ contributions to velocity
\begin{align}
 \theta&=\theta_{0}+\theta_{1}\frac{t}{\alpha}+\mathcal{O}\left(\frac{t^2}{\alpha^2}\right) ,\label{exptheta}\\ 
\omega_{\mu\nu}&=
\omega_{0\mu\nu}+\omega_{1\mu\nu}\frac{t}{\alpha}+\mathcal{O}\left(\frac{t^2}{\alpha^2}\right) ,\label{expvort}\\ 
 \sigma_{\mu\nu}&=\sigma_{0\mu\nu}+\sigma_{1\mu\nu}\frac{t}{\alpha}+\mathcal{O}\left(\frac{t^2}{\alpha^2}\right).\label{expshear}
\end{align}

It is worth noting that at linear order in curvature and time, the squared shear vanishes up to quadratic terms in time, i.e. $\sigma^2=\mathcal{O}\left(t^2/\alpha^2\right)$. In particular, this implies that
\begin{align}
\sigma_0^2=0,
\end{align}
which will be useful for the computation of the expansion. The fact that the shear squared initially vanishes is also interesting from the physical point of view, and we thoroughly discuss it in the thermodynamics section.

Following this line of reasoning and considering the goals of our study, there is one final step to be made. To have a well-defined congruence and therefore, surface~$\Sigma$ (i.e. a well-behaved system), we need to impose some constraints on the velocity field.

First, we have a constraint following directly from the definition of the velocity. We need our congruence to be timelike, and also want it to be normalized to \mbox{$g_{\mu\nu}u^\mu u^\nu=-1$}. By imposing this condition at each order in $t$, we arrive to the following results:
\begin{align} 
&u^t_{0c}=-\frac{1}{2}p_{0tt},\qquad \label{u0c}
u^t_{1c}-\frac{\alpha x_i}{r^2}u^i_{0c}=-\frac{\alpha x^i}{r^2}p_{0it},\nonumber\\
&u^t_{2c}-\frac{2\alpha x_i}{r^2}u^i_{1c}=\frac{3\alpha^2}{2r^2}p_{0tt},
\end{align}
where we have been able to identify the expression for the time component of the curved contribution at zeroth order, and to relate that same time component at first and second order to the spatial components.

In addition, we need three constraints imposed on the congruence. They lead to some extra conditions on the curved contribution to the velocity, hence refining more how it should look like.
\begin{itemize}
\item $\theta_0=0$: We need that the zeroth order contribution to the expansion (i.e. the initial value of the expansion) vanishes. Since $\theta_0=0$ in flat spacetime, any nonzero curvature-dependent contribution to it would dominate over the flat spacetime evolution of the area. The resulting surface $\Sigma$ would then no longer resemble a flat spacetime SLC, which would invalidate all our assumptions about its thermodynamics. This condition yields
\begin{equation}
\partial_i u^i_{0c}+\frac{x_i}{r^2}u^i_{0c}=\frac{1}{3}R_{ti}(p)x^i,  \label{const1}
\end{equation}
which is a differential equation on the spatial components of the curved velocity at zeroth order in time.
\item $\omega_{0\mu\nu}=0$: We want the vorticity at zeroth order in $t/\alpha$ (i.e. the initial value of the vorticity) to vanish. The reason for this constraint is that for any value of $t$, in particular, $t=0$, we want a spatial manifold with  $\Sigma$ separating the interior region from the exterior region of the SLC. Otherwise, assigning entropy to the SLC becomes problematic. From the $\omega_{0it}$ components we do not get any constraints, since its vanishing is already implied by the orthogonality of the vorticity to the velocity, $\omega_{\mu\nu}u^{\nu}=0$. The condition $\omega_{0jk}=0$ implies
\begin{equation}
\partial_j u_{0c}^k-\partial_ku_{0c}^j-u_{0c}^k\frac{x_j}{r^2}+u_{0c}^j\frac{x_k}{r^2}=\frac{1}{3}R_{jkti}\left(p\right)x^i.  \label{const2}
\end{equation}

\item $\omega_{1\mu\nu}=0$: We want the vorticity at first order in $t/\alpha$ to vanish as well, for the same reasons as the previous constraint: we want the congruence to be orthogonal to hypersurfaces.  Applying its definition (\ref{expvort}), therefore, the velocity should also satisfy
\begin{equation}
\begin{split}
\partial_j u_{1c}^i-\partial_iu_{1c}^j+\frac{\alpha}{r^2}(p_{1it}x^j-p_{1jt}x^i)=0, \label{const3}
\end{split}
\end{equation}
where, again, the $\omega_{1it}$ components do not result in any constraints (since $\omega_{\mu\nu}u^{\nu}=0$). We can check that shifting a solution of equation~\eqref{const3} by the gradient of any scalar function generates another solution (since the partial derivatives commute). Therefore, $u_{1c}^i$ can be found only up to an ambiguity given by one arbitrary function.
\end{itemize}
         
Taking all of these expressions into account, we can proceed with the two approaches in which we particularize $\theta$, $\sigma_{\mu\nu}$, and $\omega_{\mu\nu}$ for our system.

\subsubsection{Approach 1}\label{subsec:app1}

The first approach consists of the direct computation of the quantities relevant for our study, namely $u^i_c$ and~$\theta_1$. First, applying its definition, we can obtain an equation at each order in time for the acceleration. With the velocity and the acceleration, we can then compute the expressions for the expansion, shear, and vorticity of the congruence. As a result, we obtain expressions for those quantities at each order in time that depend only on the undetermined curved parts of velocity and acceleration (since the flat contributions are known). 

Solving equation \eqref{const2} to the linear order in the curvature, and imposing the condition $\theta_0=0$ discussed above, we find a unique solution for $u_{0c}^i$ (for details check Appendix~\ref{velocity constraints}) given by
\begin{equation}
u_{0c}^i=\frac{1}{3}R_{tjk}^{\ \ \ i}\left(p\right)x^jx^k.
\end{equation}
Accordingly, the most general solution for equation \eqref{const3}, linear in curvature, reads
\begin{align}
\nonumber u_{1c}^i=&-\frac{1}{6r}R_{tjtk}\left(p\right)x^jx^kx^i+\tilde{C}_1R\left(p\right)rx^i+\tilde{C}_2R_{tt}\left(p\right)rx^i \\
&+\left[\tilde{C}_3R_{jk}\left(p\right)+\tilde{C}_4R_{jtkt}\left(p\right)\right]x^j\left(2r\delta^{ki}+\frac{x^kx^i}{r}\right), \label{u^i_1c}
\end{align}
where $\tilde{C}_n,\ n=1\ldots6$ are arbitrary dimensionless constants. For a more detailed computation of these results, see Appendix~\ref{velocity constraints}. The remaining part of the velocity, $u^i_{2c}$, is not necessary and hence, we omit it.

This completes the computation of the velocity at this order in the expansion.

Taking these results into account, we now have the most general form of the curved contribution to the velocity compatible with the geometry of the stretched light cones.  Then, we can use it to compute $\theta_1$, that, together with $\sigma_{0}^2$ (which vanishes as we saw before), is the quantity of the evolution of hypersurfaces entering at this order  the thermodynamical analysis in section~\ref{section: therm}. By applying its definition and the results obtained from the constraints, we directly arrive at:
\begin{align}
\nonumber \theta_1=&\frac{2\alpha}{r^2}+\frac{1}{r^2}\partial_i\left(r^2u^i_{1c}\right)-\frac{\alpha}{3r^2}R_{ij}(p)x^ix^j
\\&-\frac{\alpha}{3}R_{tt}(p)+\frac{\alpha}{2r^2}R_{t it j}\left(p\right)x^ix^j. \label{expopt1}
\end{align}

Note that the direct computation of $\theta$ carried out here, is equivalent to the Killing identity approach to computing the changes in area of a SLC~\cite{Parikh_2018}, as we show explicitly in Appendix~\ref{Killing identity check}. Hence, we have now all the necessary ingredients for the thermodynamical analysis. However, let us first describe the other possible approach that shows entirely equivalent results, and is in fact better suited for our purposes.

\subsubsection{Approach 2}

After applying the first, direct, approach to obtain the expansion, shear, and vorticity of the congruence, we move onto the second one which uses the Raychaudhuri's equation. The way in which we proceed is to set the flat contributions to the velocity~(\ref{uflat}) and acceleration~(\ref{accelflat}) at $t=0$, i.e. $u^\mu_{0f}$  and $a^\mu_{0f}$,  as initial conditions. We also set the initial curved velocity $u^\mu_{0c}$ and the curved contribution to the acceleration up to the first order in $t/\alpha$, namely, $a^\mu_{0c}+(t/\alpha)a^\mu_{1c}$ (that can be explained as fixing the external force acting on the observer) as initial and boundary conditions of our system, for which we use the results obtained from the constraints detailed at the beginning of the section. We do so in order to ensure that the initial expansion and vorticity vanish and that the velocity is correctly normalized as we explained before. 

For the thermodynamical analysis we are interested in the expansion at first order in $t$, i.e., in $\theta_1$. Note that we already saw that $\sigma^2$ vanishes to linear order in curvature and time. Then, from Raychaudhuri's equation~(\ref{equation: evexp}), the definition of velocity~(\ref{veldef}), the definition of acceleration~(\ref{acceltime}), and the aforementioned constraints on vorticity and expansion, we can calculate the evolution of the expansion to the lowest order as
\begin{align}
\nonumber \dot{\theta} &= \frac{2}{r^2}+\nabla_\mu a^\mu_{0c} + \frac{1}{\alpha}a^t_{1c}-\frac{\alpha}{3r^2}R_{ij}(p)x^ix^j \\
&- R_{\rho\sigma}(p) u^\rho_{0f} u^\sigma_{0f} +\mathcal{O}\left(\frac{t}{\alpha}\right), \label{dot theta}
\end{align}
where we have used that the divergence of the acceleration equals
\begin{align}
\nabla_\mu a^\mu= \frac{2}{r^2}+\nabla_\mu a^\mu_{0c}+\frac{1}{\alpha}a^t_{1c}-\frac{1}{3r^2}R_{ki}(p)x^ix^k+\mathcal{O}\left(\frac{t}{\alpha}\right), \label{nablamuamu}
\end{align}
where the term $2/r^2$ is just the flat spacetime contribution.

On the other hand, since the initial conditions ensure $\theta_0=0$, we find, to leading order in time, that the expansion~\eqref{exptheta} of $\theta$ yields
\begin{equation}
\dot{\theta}=\frac{\dot{t}}{\alpha}\theta_1+\mathcal{O}\left(\frac{t}{\alpha}\right),
\end{equation}
where, using equation~\eqref{u0c}, we can write
\begin{equation}
\dot{t}=u^{\mu}\nabla_{\mu}t=u^{t}=1-\frac{1}{6}R_{jtkt}(p)x^jx^k+\mathcal{O}\left(\frac{t}{\alpha}\right).
\end{equation}
Thence, it holds
\begin{equation}
\dot{\theta}=\frac{1}{\alpha}\theta_1\left[1-\frac{1}{6}R_{titj}(p)x^i x^j\right]+\mathcal{O}\left(\frac{t}{\alpha}\right).
\end{equation}
This expression, together with equation~\eqref{dot theta} allow us to solve for $\theta_1$:
\begin{align}
\nonumber \theta_1=&\frac{2\alpha}{r^2}\left(1+\frac{1}{6}R_{t it j}\left(p\right)x^ix^j\right)+\alpha\nabla_\mu a^\mu_{0c}+a^t_{1c} \\
&-\alpha R_{\rho\sigma}(p) u^\rho_{0f} u^\sigma_{0f}-\frac{\alpha}{3r^2}R_{ij}(p)x^ix^j. \label{expopt2}
\end{align}

\subsubsection{Comparison between the approaches}\label{sec: comparison}

To conclude, we compare both approaches and see whether their application (e.g. \cite{Chirco_2010} and \cite{Parikh_2018}) result in the same physics or not. Our intuition would tell us that both of them should be completely equivalent or else one of them should fail to reproduce the Einstein's equations when applying the thermodynamic formalism in the next section. In order to compare the results from both of them, we focus on the expansion, since it is the quantity used for the thermodynamical analysis. The first approach yielded the result given by the identity~\eqref{expopt1} and the second one led to identity \eqref{expopt2}.

In the first approach, we have initially three free functions $u^i_{1c}$, corresponding to the three spatial components of the curved contribution to the velocity at first order in time. However, the constraint equation~\eqref{const3} reduces the ambiguity in $u^i_{1c}$ to the gradient of some scalar function, i.e., to one free function. On the other hand, we also see that the second approach relies as well on a free function, $a^t_{1c}$: the time component of the curved contribution to the acceleration, at first order in $t$. Therefore, in terms of the number of free functions, both approaches are equivalent.

A direct way to see that both approaches are completely equivalent is to compute the curved components of the acceleration using its definition in terms of the velocities, replace them in equation~\eqref{expopt2} and check that the result for the expansion coincides with the expression of the first approach~\eqref{expopt1}.  We checked that, in this case, equations~\eqref{expopt1} and~\eqref{expopt2} both yield the same result for the expansion. That is what we would expect to happen. From now on, we use \eqref{expopt2} as the expression for $\theta_1$, since it can be physically interpreted more directly.

\subsection{Area change on the SLC}
\label{sub:area}

Obtaining the expansion allows us to directly compute the change of the area of the SLC, which is crucial for the definition of entropy in the following thermodynamics analysis.

The change of area along the congruence obeys
\begin{equation}
\delta A=\int_\Sigma \theta \text{d}t\text{d}A ,\label{raycharea}
\end{equation}
where $\theta$ is the expansion of the generators of the SLC. This equation holds since the expansion is the rate of change of the cross sectional area, orthogonal to the velocity vector $u^{\mu}$~\cite{Raych}. In section \ref{section: geom}, we imposed that the expansion vanishes at $t=0$. Moving from the cross sectional surface $t=0$ along the congruence defined by $u^\mu$, the evolution of the expansion $\theta$ is given by an expansion in powers of $t$ around it. This expansion is precisely given in equation~\eqref{exptheta}, where we have imposed that $\theta_0=0$ and obtained an expression for $\theta_1$ (we have shown that we can equivalently proceed in two ways, either considering only a time expansion, or applying Raychaudhuri's equation). As we mentioned at the end of the subsection \ref{sec: comparison}, we work with the result from Raychaudhuri's approach given by equation \eqref{expopt2}. In consequence, equation~\eqref{raycharea} becomes
\begin{align}
\nonumber \delta A=&\int_\Sigma \bigg[\frac{2}{r^2}\left(1+\frac{1}{6}R_{t it j}\left(p\right)x^ix^j\right)+\nabla_\mu a^\mu_{0c}+a^t_{1c} \\
&-R_{tt}(p)-\frac{1}{3r^2}R_{ij}(p)x^ix^j\bigg]t\hspace{0.1cm}\text{d}t\text{d}A+\mathcal{O}\left(\frac{\epsilon^3}{\alpha^3}\right),\label{raytherm}
\end{align}
where the subleading $\mathcal{O}\left(\epsilon^3/\alpha^3\right)$ terms do not affect our analysis and they are going to be systematically discarded from now on. The first contribution \mbox{$2/r^2=\partial_{\mu}a^{\mu}_{0f}+a^{t}_{1f}$} is unique in that it is associated with the observer's acceleration and does not vanish in flat spacetime. In particular, the first term of the previous equations gives the change of area in flat spacetime
\begin{equation}
\delta A_{f}=\int_\Sigma\frac{2}{r^2}t\text{d}t\text{d}A=4\pi\epsilon^2. \label{areaflat}
\end{equation}

\section{Energy balance}\label{section: energy}

We are interested in obtaining the equations governing the dynamics of a generic spacetime from thermodynamics. In order to achieve that, we introduce thermodynamic quantities relevant for the local energy balance in the SLC. A crucial role is played by the test observer moving with velocity $u^{\mu}$, corresponding to approximately constant acceleration, along the SLC. This observer measures both the Unruh temperature and the energy flux across the SLC. Moreover, as we show, the properties of the observer actively enter the energy balance, since the work required to maintain their constant acceleration cannot be neglected.

We study the energy balance by means of the heat flux $\delta Q$ through $\Sigma$, in the form of exchanges between our system and its surroundings. However, that is a concept introduced in the frame of nonrelativistic thermodynamics. Its nature in the relativistic context is rather subtle. In our case, we can associate heat $\delta Q$ with (part of) the flux of matter and energy (i.e., with $T_{\mu\nu}$, the energy-momentum tensor) through the SLC (see FIG.~\ref{fig:Fluxx}). In general, $T_{\mu\nu}$ can have a fully randomized part corresponding to the heat flux, together with work terms associated, e.g. with pressure or with electric charge. The correct division of $T_{\mu\nu}$ into heat flux and work terms is essential in order to correctly derive and fully understand the resulting Einstein's equations. Thus, in the presence of work terms, $\delta Q$ is defined by
\begin{equation}
\delta Q=\delta E+\delta W, \label{deltaQEW}
\end{equation}
where $\delta E$ is the energy-momentum flux through $\Sigma$ and $ \delta W$ is the work done on the observer (hence the positive sign). 

More precisely, $\delta E$ is the total flux (consisting of the background and the contribution of the accelerating observer together) along the velocity field $u^\mu$, integrated over the SLC, $\Sigma$, i.e.,
\begin{equation}
\delta E=-\int_\Sigma T_{\mu\nu}u^\mu\text{d}\Sigma^\nu, \label{heat}
\end{equation}
with $d\Sigma^\nu$ being the volume element of the SLC and the minus sign necessary for the flux to be future-directed. To define $d\Sigma^\nu$, we need to specify an outward-pointing, unit vector normal to $\Sigma$. This spacelike vector is proportional to the acceleration vector $a^\mu$, namely, $\alpha a^\mu$, which has unit norm. This allows us to rewrite equation (\ref{heat}) as
\begin{equation}
\delta E=-\alpha\int_\Sigma T_{\mu\nu}u^\mu a^\nu\text{d}t\text{d}A.  \label{deltaq}
\end{equation}

In order to define the work done on the system, we choose the background energy-momentum to be pure heat and, hence, by the balance of the first law of thermodynamics~\cite{Parikh:2020}, not contributing to the work done on the system. Therefore, the energy and work terms balancing the summed up heat terms, are defined in terms of the observer's system. In analogy to standard thermodynamics, we treat the exterior of the SLC as a heat bath interacting with the system under study, i.e., with the SLC. We can then conclude that the work term must be entirely associated with the accelerating observer. We may interpret it as the work done on (or by) the observer to maintain it at constant acceleration.

Working for the time being in flat spacetime, let us now estimate the work $\delta W$ necessary to maintain the observer's acceleration. A uniformly accelerating observer needs to exert power to maintain their acceleration. In general, this power is the product of the time component of the observer's acceleration and their total rest energy. In the nonrelativistic limit, this formula reduces to the standard expression $P=\vec{F}\cdot\vec{v}$. 

The time component of the acceleration of the observer moving along the SLC in flat spacetime equals $a^t=a^t_{\text{f}}=t/\alpha^2$. Let us now estimate their rest energy. As we discussed in subsection~\ref{sub:SLC}, the observer thermalizes to the Unruh temperature $T_{\text{U}}$ proportional to the modulus of their acceleration, $1/\alpha$. Then, by the zeroth law of thermodynamics, the observer's temperature also becomes $T_{\text{U}}$, and, in their rest frame, they acquire an internal energy $U$ proportional to $T_{\text{U}}$, and to the number of internal degrees of freedom that can be thermally excited. Since $T_{\text{U}}$ is in our case very large, we can safely assume that $U$ becomes far larger than the rest mass of the observer. The total rest energy of the observer can then be approximated by $U$, which equals
\begin{equation}
U=\int \nu T_{\text{U}}\text{d}A,
\end{equation}
where the integral is carried out over the spacelike cross-section of the SLC and $\nu$ denotes the areal density of the internal degrees of freedom (which we fix in the next section). The power $P$ necessary to maintain the observer's acceleration then equals
\begin{equation}
\label{power}
P=\frac{\text{d}}{\text{d}\tau}\left(\delta W\right)=Uu^{\nu}\nabla_{\nu}u^t=a^t\int \nu T_{\text{U}}\text{d}A.
\end{equation}
Integration in proper time then yields the total work, to the leading order in $\epsilon$,
\begin{equation}
\delta W=\int_{\Sigma}\frac{t}{\alpha^2}\nu T_{\text{U}}\text{d}A \text{d}t+\mathcal{O}\left(\frac{\epsilon^3}{\alpha^3}\right),
\end{equation}
where, up to higher order correction in $\epsilon$, we were able to replace the integration in the proper time with integration in the coordinate time $t$.

In a generic curved spacetime, the expression for the work necessary to accelerate the observer changes due to the gravitational forces acting on it. However, equation~\eqref{power} for the power remains valid and its integral in proper time yields
\begin{equation}
\label{work curved}
\delta W=\int_\Sigma \nu T_{\text{U}}\frac{t}{\alpha^2}\left[1+\frac{1}{6}R_{titj}\left(p\right)x^ix^j\right]\text{d}t\text{d}A+\mathcal{O}\left(\frac{\epsilon^3}{\alpha^3}\right).
\end{equation}
This is the expression for work we consider in the next section.

\section{Entropy balance}\label{section: therm}

Having studied the geometrical properties of our system, and the energy balance governing it, we are now ready to finally proceed with the thermodynamical analysis. In order to do so, we use the previously defined SLC. Through Clausius' equilibrium relation, we associate its geometrical properties to the thermodynamical ones
\begin{equation}
\delta S=\frac{\delta Q}{T_{\text{U}}}, \label{clausius}
\end{equation}
where $\delta S$ denotes the change in the entropy of the SLC, $\delta Q$ is the heat flux crossing the SLC introduced in the previous section, and $T_{\text{U}}$ is the Unruh temperature given by equation~\eqref{Unruh}, as seen by our accelerating observer defining the SLC.

We also state the relation between the entropy $S$ and the area $A$ of the system, as follows
\begin{equation}
\delta S=\eta \delta A, \label{antrarea}
\end{equation}
with $\eta$ being some constant with the units of inverse area, so that the entropy is dimensionless. Via entanglement entropy arguments in QFT \cite{SA,SAA,Solodukhin:2011}, it is safe to assume that this relation holds for our system, as long as the radius of curvature of our spacetime is much larger than Planck's length \cite{Jacobson_2016}. Later on, we discuss the relation of $\eta$ with the Newton's gravitational constant $G$ (along the same lines as in previous works on the subject~\cite{Jacobson_2016,Alonso:2024}).

Therefore, combining equations~\eqref{clausius} and~\eqref{antrarea}, we obtain a direct relation between the heat exchanged by our system, and its area, which encodes information about the geometry, as
\begin{equation}
\eta \delta A=\frac{\delta Q}{T_{\text{U}}}. \label{clausiusss}
\end{equation}
By relating the change of area with the spacetime curvature as we have done in section~\ref{section: geom}, we are able to connect the curvature to $\delta Q$ through equation \eqref{clausiusss}. 

Applying equations~\eqref{raytherm},~\eqref{deltaQEW}, and~\eqref{deltaq} to the Clausius relation~\eqref{clausiusss} yields
\begin{align}
&\int_\Sigma \bigg[\frac{2}{r^2}\left(1+\frac{1}{6}R_{t it j}\left(p\right)x^ix^j\right)+\nabla_\mu a^\mu_{0c}+a^t_{1c} \notag\\
&-R_{tt}(p)-\frac{1}{3r^2}R_{ij}(p)x^ix^j\bigg]t\hspace{0.1cm}\text{d}t\text{d}A \notag\\
\hspace{0.2cm}&=-\frac{1}{\eta T_{\text{U}}}\int_\Sigma  \alpha
\hspace{0.1cm} T_{\mu\nu}u^\mu a^\nu\text{d}t\text{d}A+\frac{1}{\eta T_{\text{U}}}\delta W.  \label{equivalqt}
\end{align}

We have previously shown that the work necessary to accelerate the observer obeys equation~\eqref{work curved}, where the areal density $\nu$ remains undetermined. To fix it, we take a small detour discussing the content of equation~\eqref{equivalqt} in flat spacetime. It reads
\begin{align}
\nonumber \int_\Sigma\frac{2}{r^2}t\text{d}t\text{d}A =&-\frac{1}{\eta T_{\text{U}}}\int_\Sigma \alpha T_{\mu\nu}u_f^\mu a_f^\nu\text{d}t\text{d}A \\
&+\frac{1}{\eta T_{\text{U}}}\int_{\Sigma}\frac{t}{\alpha^2}\nu T_{\text{U}}\text{d}A \text{d}t,
\end{align}
where $T_{\mu\nu}$ corresponds solely to the energy-momentum associated with the observer. To integrate this expression, we approximate both $T_{\mu\nu}$ and $\nu$ by their values at $p$ (the other terms in their Taylor expansion are suppressed by higher powers of $\epsilon/\alpha$), so that they are constant as far as the integration is concerned. For integrating $u^{\mu}_fa^{\nu}_f$, we use that the integral of $x^i$ over a sphere vanishes and $x^ix^j$ integrates to $4\pi\delta^{ij}/3$. We finally obtain
\begin{align}
\nonumber -4\pi\epsilon^2 g_{tt}\left(p\right)=&-\frac{4\hbar}{3\eta}\epsilon^2\left[T_{tt}\left(p\right)-\frac{1}{4}T\left(p\right)g_{tt}\left(p\right)\right] \\
&-\frac{2\pi}{\eta}\epsilon^2\nu\left(p\right)g_{tt}\left(p\right),
\label{flat Clausius integrated}
\end{align}
where we used that $g_{tt}\left(p\right)=-1$ to write the entire expression as a time-time component of some tensor. Equation~\eqref{flat Clausius integrated} actually holds as a tensorial identity. The proof lies in the fact that, given any rank 2 tensor $X_{\mu\nu}$, we can express its temporal component as follows: \mbox{$X_{tt}=X_{\mu\nu}\delta^\mu_t\delta^\nu_t$}. Since \mbox{$u^\mu_{0f}=\delta^\mu_t$}, we can rewrite this expression as \mbox{$X_{tt}=X_{\mu\nu}u^\mu_{0f}u^\nu_{0f}$}. Let us now choose a particular tensor $X_{\mu\nu}$ such that equation \eqref{flat Clausius integrated} can be rewritten as $X_{tt}=0$, or, equivalently,
\begin{align}
X_{\mu\nu}u^\mu_{0f}u^\nu_{0f}=0.
\end{align}
This equation is true for any choice of a unit, timelike, future-pointing vector $u^\mu_{0f}$ at $p$. Therefore, it holds that $X_{\mu\nu}=0$ (see Appendix B of reference~\cite{Alonso:2024} for the explicit proof of this statement). Equation~\eqref{flat Clausius integrated} then implies
\begin{align}
\nonumber -4\pi\epsilon^2 g_{\mu\nu}\left(p\right)=&-\frac{4\hbar}{3\eta}\epsilon^2\left[T_{\mu\nu}\left(p\right)-\frac{1}{4}T\left(p\right)g_{\mu\nu}\left(p\right)\right] \\
&-\frac{2\pi}{\eta}\epsilon^2\nu\left(p\right)g_{\mu\nu}\left(p\right). \label{flat Clausius tensor}
\end{align}
All the tensors besides $T_{\mu\nu}\left(p\right)$ are proportional to $g_{\mu\nu}\left(p\right)$. Thence, to satisfy equation~\eqref{flat Clausius tensor}, it must hold
\begin{equation}
T_{\mu\nu}\left(p\right)=\frac{1}{4}T\left(p\right)g_{\mu\nu}\left(p\right),
\end{equation}
where the trace $T$ of the energy-momentum tensor is an arbitrary function. The first term on the right hand side of equation~\eqref{flat Clausius tensor} then vanishes and we obtain
\begin{align}
-4\pi\epsilon^2 g_{\mu\nu}\left(p\right)=-\frac{2\pi}{\eta}\epsilon^2\nu\left(p\right)g_{\mu\nu}\left(p\right),
\end{align}
implying
\begin{align}
\label{nu}
\nu\left(p\right)=2\eta,
\end{align}
up to higher order corrections in $\epsilon/\alpha$. Let us note that the first term of the integrand in equation~\eqref{work curved} precisely cancels out with the flat spacetime contribution to the change of area.

Once we have fixed the value of $\nu(p)$, we can proceed to integrate equation \eqref{equivalqt}. On the right hand side, we can expand $T_{\mu\nu}$ around $p$, keeping only the leading order constant term $T_{\mu\nu}(p)$. Moreover, any terms of the form $T_{\mu\nu}$ times Riemann tensor are effectively quadratic in curvature, and, being interested only in the contributions linear in curvature, we can discard them. Hence, we only need to integrate $u^{\mu}_fa^{\nu}_f$, as we did previously in the flat spacetime case. We further discard any contributions of the order $\mathcal{O}\left(\epsilon^3/\alpha^3\right)$.

Regarding the left hand side, we first evaluate the acceleration terms that, using the results of subsection~\ref{sec:omsigex}, read
\begin{align}
\nonumber &\nabla_\mu a^\mu_{0c}+a^t_{1c}\\  &=\left(-\frac{2}{3}+6\tilde{C}_4\right)\frac{x^ix^j}{r^2}R_{titj}\left(p\right)+\left(6\tilde{C}_1+2\tilde{C}_3\right)R\left(p\right) \\
&+\left(\frac{2}{3}+6\tilde{C}_2+2\tilde{C}_3+2\tilde{C}_4\right)R_{tt}\left(p\right)+6\tilde{C}_3\frac{x^ix^j}{r^2}R_{ij}\left(p\right).\nonumber
\end{align}
All the curvature tensors are evaluated at $p$ and, hence, they are constant. The integration is then simple, using the same approach as for the right hand side.

In total, the integration yields
\begin{align}
C_1R_{tt}\left(p\right)+C_2R\left(p\right)g_{tt}\left(p\right)=&\frac{24\pi}{\eta \hbar}\left[T_{tt}\left(p\right)-\frac{1}{4}T\left(p\right)g_{tt}\left(p\right)\right],\label{thermsol}
\end{align}
where we used that $g_{tt}\left(p\right)=-1$, and also multiplied the equation by the factor $9/(2\pi\epsilon^2\alpha^2\eta)$. We have used the definition for temperature given in equation \eqref{Unruh}, $R$ is the Ricci scalar, and $T=T^\mu_{\ \mu}$ is the trace of the energy-momentum tensor. The undetermined numbers $C_1$, $C_2$ correspond to the ambiguities in the constrained form of $u^{i}_{1c}$, and, in terms of the parameters introduced in equation~\eqref{u^i_1c}, they read
\begin{align}
C_1&=6-54\tilde{C}_2-36\tilde{C}_3-36\tilde{C}_4, \label{C1} \\
C_2&=-1+54\tilde{C}_1+36\tilde{C}_3. \label{C2}
\end{align}

From equation~\eqref{thermsol}, we can obtain a tensorial identity that will end up providing us with Einstein's equations, following the same reasoning as in the flat spacetime case above. In terms of Einstein's tensor \mbox{$G^{\mu\nu}$}, this identity can be expressed as
\begin{align}
\nonumber &C_1G_{\mu\nu}\left(p\right)+\left(\frac{1}{2}C_1+C_2\right)R\left(p\right)g_{\mu\nu}\left(p\right) \\
&=\frac{24\pi}{\eta \hbar}\left[T_{\mu\nu}\left(p\right)-\frac{1}{4}T\left(p\right)g_{\mu\nu}\left(p\right)\right], \label{eseqs}
\end{align}
which are tensorial equations linking spacetime curvature at the point $p$ with the matter content of the spacetime. The strong equivalence principle then guarantees that these equations do not hold only at $p$, but throughout the spacetime~\cite{Jacobson_1995,Chirco_2010}.

\begin{figure}[tbp]
\centering
\includegraphics[width=1\columnwidth,origin=c,trim={5.95cm 20.1cm 5.8cm 1.8cm},clip]{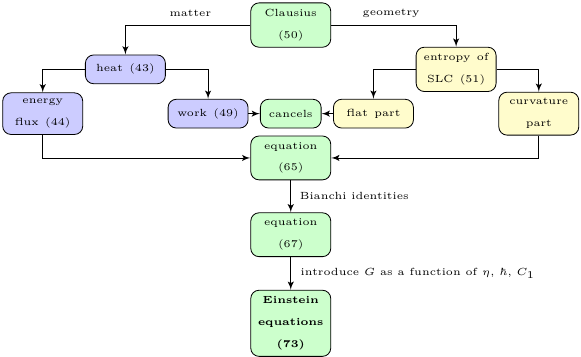}
\caption{ \label{diagram} Flow diagram of the derivation of Einstein equations from the Clausius equilibrium relation. Blue boxes in the diagram correspond to matter contributions, yellow boxes to geometry contributions, and green boxes to their combination.}
\end{figure}

Since the energy-momentum tensor captures the total matter content of the spacetime, including that one associated with the accelerating observer, it should be locally conserved, i.e., $\nabla_\mu T^{\mu\nu}=0$\footnote{Let us note that in the context of unimodular gravity one could impose instead the less restrictive condition $\tilde{\nabla}_{\nu}\tilde{T}_{\mu}^{\;\:\nu}=\tilde{\nabla}_{\mu}\mathcal{J}$~\cite{Carballo:2022}.}. The significance of this observation lies beyond the scope of this paper, and has been analyzed elsewhere~\cite{Alonso:2024}. Then, the remaining terms in equation \eqref{eseqs}, must also have vanishing divergence. We know that $\nabla_\mu G^{\mu\nu}=0$ and, therefore, taking the divergence of \eqref{eseqs} gives an equation for $\nabla_\mu T$ and $\nabla_\mu R$ which, upon integration, yields
\begin{align}
T=\frac{\eta \hbar}{6\pi}\left(-\frac{1}{2}C_1-C_2\right)R+\Tilde{\Lambda},\label{TR}
\end{align}
with $\Tilde{\Lambda}$ some integration constant. Therefore, taking the relation given by \eqref{TR}, and introducing it into \eqref{eseqs}, we see that
\begin{align}
C_1G_{\mu\nu}+\frac{6\pi}{\eta \hbar}\Tilde{\Lambda}g_{\mu\nu}=\frac{24\pi}{\eta \hbar}T_{\mu\nu}. \label{ecasi}
\end{align}
This result already looks like Einstein's equations, up to three unknown constants: $C_1$, $\eta$ and $\Tilde{\Lambda}$. We are going to see that we are not able to entirely fix them, but we can constrain their values. For that purpose, we first impose that the non-relativistic limit of the equations, should recover Newtonian dynamics. This limit provides a relation between the constants $C_1$ and $\eta$, and the Newton gravitational constant $G$, which reads
\begin{align}
8\pi G=\frac{24\pi}{\eta \hbar}\frac{1}{C_1}.
\end{align}
The relation \eqref{TR} between $T$ and $R$ must also be consistent with the non-relativistic limit, fixing the proportionality factor to $-1/\left(8\pi G\right)$. We can then build a system of equations given by these two conditions, that leads us to the following relation between $C_1$, $C_2$, and $\eta$:
\begin{align}
C_2=-\frac{3}{4}C_1,\qquad
\eta=\frac{1}{4G\hbar}\frac{12}{C_1}. \label{cts}
\end{align}
The expression for $\eta$ differs between the timelike stretched light cone and null strict light cones. For the latter, one obtains the fixed value \mbox{$\eta=1/\left(4G\hbar\right)$}~\cite{Hawking,Jacobson:2019}. More generally, for any null surface the recovery of Einstein equations from thermodynamics requires \mbox{$\eta=1/\left(4G\hbar\right)$}~\cite{Jacobson_1995,Jacobson_2016}. Nevertheless, as already stated, we are working with SLCs, i.e. timelike surfaces. In this case, we are not able to fix the value of the proportionality constant between $S$ and $A$. The value $C_1=12$ reproducing the Bekenstein entropy represents just one of the possibilities. In principle, any $C_1>0$ leads to positive entropy and is admissible.

In the context of physics of horizonless black hole mimickers, it has suggested that entropy of timelike hypersurfaces near the null horizon is close to the Bekenstein value~\cite{Abreu:2010,Barcelo:2011a,Barcelo:2011b,Mathur:2023,Mathur:2024}, i.e.,
\begin{equation}
\eta=\frac{1}{4G\hbar}+f(\varepsilon).   
\end{equation}
where $\varepsilon$ is a small parameter such that it vanishes in the limit of the timelike hypersurface approaching the null one. Assuming the same result applies to SLCs, we can expect that $C_1=12+\varepsilon$, where $\lim_{\alpha\to0}\varepsilon=0$ (as the limit $\alpha\to0$ corresponds to the light cone). Then, we have
\begin{equation}
\eta\approx\frac{1}{4G\hbar}+\frac{\varepsilon}{48G\hbar}+\mathcal{O}\left(\varepsilon^2\right).
\end{equation}

Finally,
\begin{align}
\Lambda=2\pi G\Tilde{\Lambda}, \label{lambda}
\end{align}
can be identified with the cosmological constant. Consequently, plugging the expressions given in equations \eqref{cts} and \eqref{lambda}, into equation \eqref{ecasi}, we finally arrive at
\begin{align}
G_{\mu\nu}+\Lambda g_{\mu\nu}=8\pi G T_{\mu\nu}, \label{Einstein}
\end{align}
that are the sought for Einstein's equations\footnote{On the same grounds as in the previous work of three of the authors~\cite{Alonso:2024}, it can be argued that the resulting gravitational dynamics are consistent with Weyl transverse gravity rather than with general relativity. Since this difference is unimportant for the purposes of the present paper, we do not discuss it here.}. For convenience, we summarize the basic logic of our derivation of equations~\eqref{Einstein} from equilibrium thermodynamics in the flow diagram~\ref{diagram}.

\section{Conclusions}\label{section: conclusions}

This paper explores the relation between thermodynamics of SLCs and the gravitational dynamics from a new angle. Previous works on the topic stressed the importance of non-equilibrium entropy production for thermodynamics of local causal horizons. This entropy production has been said to be sourced by shear and by the flat spacetime expansion of the SLC. Herein, we have computed in detail the expansion, shear, and vorticity for the SLCs, separating flat and curved contributions, to introduce them into an energy and entropy thermodynamic balance, carefully identifying the work term, and deriving the equations governing the gravitational dynamics. We show here that the shear scalar squared vanishes to the required order in time and curvature. Moreover, we interpret the flat spacetime contribution, being associated with the acceleration, as the work necessary to accelerate the observer. As a result, in contrast to the existing literature~\cite{Chirco_2010,Parikh_2018}, we actually show how one can describe the behaviour of SLCs fully in terms of equilibrium thermodynamics. 

On the more technical level, we explore the freedom present in defining SLCs in a generic curved spacetime. Specifically, we focused on the expansion, shear, and vorticity of the accelerating congruence, whose timelike slice forms the SLC. These properties were studied under two different approaches: one relying on an explicit construction of the velocity vector field, and the other based on Raychaudhuri's equation. By imposing the conditions of vanishing vorticity and initial expansion, we describe the most general SLC suitable for thermodynamic considerations. We also prove the complete equivalence between our method and the previously employed technique relying on the approximate Killing identity, to compute the area of the SLC~\cite{Parikh_2018}.

Furthermore, by applying Clausius' relation alongside the entropy-area law motivated by entanglement entropy arguments in QFT \cite{SA,SAA}, we derive Einstein's field equations. This is ultimately achieved by relating the results obtained in the geometrical study of the system, to the thermodynamical analysis of the heat fluxes present in the spacetime. However, an important aspect to bear in mind is that, as opposed to prior research, our resulting dynamical equations rely on the value of a free parameter to fix the proportionality constant between the variation of the entropy and that of the area. This previously unnoticed feature appears to be characteristic for thermodynamics of timelike surfaces.

Our work focused on the leading order contributions to the Clausius equilibrium relation in powers of the SLC's size parameter $\alpha$ and the short time scale $\epsilon$. Since we consider both $\alpha$ and $\epsilon$ to be much smaller than the local curvature length scale, they certainly represent only small corrections to the leading order terms we studied. Nevertheless, a natural question is whether these subleading contributions can encode corrections to the Einstein's equations. Two of us have previously studied this question in the vacuum setting~\cite{Alonso:2025}. We found that, since these correction terms include four or more derivatives of the metric, for dimensional reasons they would appear in the gravitational equations multiplied by suitable powers of $\alpha$ and/or $\epsilon$. However, as we discussed, both of these length scales are arbitrary (within a given range) in our treatment. Hence, the gravitational dynamics cannot depend on them. To study the corrections to the Einstein's equations in this way, one would need to introduce a physically motivated length scale (e.g. in the way we explored here~\cite{Alonso:2025}).

Overall, we provide a unified and equilibrium-based framework that connects thermodynamic laws with gravitational dynamics, specifically through the use of stretched light cones. This research opens the doors to new questions: What is the precise definition of work and heat in the context of curved spacetimes? Is it meaningful to associate entropy with an observer-dependent timelike surface, despite its apparent arbitrariness? How can observer dependence shape our understanding of spacetime entropy? We hope to address these questions in future research.

\section*{Acknowledgments}

AA-S is funded by the Deutsche Forschungsgemeinschaft (DFG, German Research Foundation) — Project ID 51673086. ML is supported by the DIAS Post-Doctoral Scholarship in Theoretical Physics 2024. AA-S and LJG acknowledge support through Grant  No.  PID2023-149018NB-C44 (funded by MCIN/AEI/10.13039/501100011033).

\appendix

\section{Solving the velocity constraints}
\label{velocity constraints}

In section \ref{sec:omsigex}, we introduced the constraints that are to be applied in our system. In the first approach to computing $\theta$, $\sigma_{\mu\nu}$ and $\omega_{\mu\nu}$, we have shown the most general results to the equations imposed by said constraints, under the formalism applied in that same approach. In this Appendix, we aim to explicitly show how to obtain such results.

The first ones that we will deal with, are the equations~\eqref{const2} obtained from imposing $\omega_{0\mu\nu}=0$. These equations have a general solution
\begin{align}
\nonumber &u_{0c}^i=\frac{1}{3}R_{tjk}^{\ \ \ i}\left(p\right)x^jx^k+\mathcal{K}^{i}_{\ i_1\dots i_m}\left(p\right)rx^{i_1}...x^{i_m} \\
\nonumber &+\left[\frac{n}{q+2}\mathcal{L}^{i}_{\ i_1\dots i_{n-1}}\left(p\right)r^2-\mathcal{L}_{i_1\dots i_n}\left(p\right)x^{i_n}x^{i}\right] \\
&\times r^{q}x^{i_1}...x^{i_{n-1}},
\end{align}
where $m$, $n$, $q$ are arbitrary natural numbers and $\mathcal{K}^{i}_{\ i_1\dots i_m}$, $\mathcal{L}^{i}_{\ i_1\dots i_n}$ arbitrary tensors symmetric in all indices and constructed from the Riemann tensor, the covariant derivative and the metric. The first term represents the particular solution and the latter are the homogeneous solutions of equation~\eqref{const2}. We stress that the tensors evaluated at $p$ behave as constant under derivatives.

In this work, we focus on corrections to $u^{\mu}$ linear in the spacetime curvature. Then, we have a simpler solution
\begin{align}
\nonumber u_{0c}^i=&\frac{1}{3}R_{tjk}^{\ \ \ i}\left(p\right)x^jx^k+\tilde{D}_1R_{j}^{\ i}\left(p\right)rx^{j} \\
\nonumber &+\tilde{D}_2R^{i}_{\ tjt}\left(p\right)rx^j+\tilde{D}_3R_{jk}\left(p\right)\frac{1}{r}x^{j}x^{k}x^{i} \\
&+\tilde{D}_4R_{jtkt}\left(p\right)\frac{1}{r}x^jx^kx^i,
\end{align}
where $\tilde{D}_n,\ n=1\ldots4$ are arbitrary dimensionless constants.

If we also impose condition~\eqref{const1} implied by vanishing of the initial expansion $\theta_0$, we then obtain $\tilde{D}_1=\tilde{D}_2=\tilde{D}_3=\tilde{D}_4=0$. This leaves us with the unique solution for $u_{0c}^i$ given by
\begin{equation}
u_{0c}^i=\frac{1}{3}R_{tjk}^{\ \ \ i}\left(p\right)x^jx^k.
\end{equation}

The other set of equations that we will work with in this Appendix, is the set of equations obtained from imposing $\omega_{1\mu\nu}=0$; equations \eqref{const3}. The most general solution of this equation reads
\begin{align}
\nonumber u_{1c}^i&=-\frac{1}{6r}R_{tjtk}x^jx^kx^i+\mathcal{M}^{i}_{\ i_1\dots i_m}x^{i_1}\dots x^{i_m} \\
\nonumber &+\mathcal{N}r^{a}x^{i}+\left(n\mathcal{P}^{i}_{\ i_1\dots i_{n-1}}r^2+b\mathcal{P}_{i_1\dots i_n}x^nx^i\right) \\&\times r^{b-2}x^{i_1}\dots x^{i_{n-1}},
\end{align}
where $a$, $b$ are arbitrary real numbers, $m$, $n$ arbitrary natural numbers and $\mathcal{M}^{i}_{\ i_1\dots i_m}$, $\mathcal{N}$, $\mathcal{P}_{i_1\dots i_n}$ are again arbitrary tensors symmetric in all indices and vanishing in flat spacetime. As the structure of equation~\eqref{const3} suggests, $u^i_{1c}$ has the form of a particular solution plus a gradient of an arbitrary scalar function
\begin{align}
\nonumber u_{1c}^i=&-\frac{1}{6r}R_{tjtk}x^jx^kx^i+\partial_{i}\Big(\frac{1}{m+1}\mathcal{M}_{\ i_1\dots i_{m+1}}x^{i_1}\dots x^{i_{m+1}} \\
&+\frac{1}{a+2}\mathcal{N}r^{a+2}+\mathcal{P}_{\ i_1\dots i_n}r^bx^{i_1}\dots x^{i_n}\Big),
\end{align}

The most general solution linear in curvature reads
\begin{align}
\nonumber u_{1c}^i=&-\frac{1}{6r}R_{tjtk}\left(p\right)x^jx^kx^i+\tilde{C}_1R\left(p\right)rx^i+\tilde{C}_2R_{tt}\left(p\right)rx^i \\
&+\left[\tilde{C}_3R_{jk}\left(p\right)+\tilde{C}_4R_{jtkt}\left(p\right)\right]x^j\left(2r\delta^{ki}+\frac{x^kx^i}{r}\right),
\end{align}
where $\tilde{C}_n,\ n=1\ldots6$ are arbitrary dimensionless constants.

\section{Equivalence of the expansion and Killing identity calculations}
\label{Killing identity check}

The seminal derivation of the Einstein's equations from thermodynamics of SLCs, studied the changes in their area through the use of an approximate Killing identity applied to velocity $u^{\mu}$~\cite{Parikh_2018}. Using that the change in area between times $t=0$ and $t=\epsilon$ equals\footnote{The calculation is originally carried out with the boost vector $\xi^{\mu}$. However, the authors treat $\xi^{\mu}$ and $u^{\mu}$ as being directly proportional to each other, making our expression with $u^{\mu}$ equivalent to the one they use~\cite{Parikh_2018}.}
\begin{equation}
\label{area xi}
\delta A=\frac{1}{2}\int_{\Sigma}\left(\nabla_{\nu}\nabla^{\nu}u^{\mu}-\nabla_{\nu}\nabla^{\mu}u^{\nu}\right)\alpha n_{\mu}\text{d}t\text{d}A,
\end{equation}
the approximate Killing identity yields
\begin{equation}
\delta A=\int_{\Sigma}\left(R_{\mu\nu}u^{\nu}+f_{\ (\mu\nu)}^{\nu}\right)\alpha n^{\mu}\text{d}t\text{d}A,
\end{equation}
where tensor $f_{\ (\mu\nu)}^{\nu}$ accounts for the fact that $u^{\mu}$ is not a Killing vector. The term proportional to the Ricci tensor then enters the Einstein's equations, whereas $f_{\ (\mu\nu)}^{\nu}$ gives rise to additional entropy production that is, to the leading order, of the same form as the $\nabla_{\mu}a^{\mu}$ term in our Raychaudhuri equation analysis. While the seminal paper interprets it as irreversible entropy production, we show that it is in fact compensated by the work necessary to accelerate the observer.

It is far from obvious that equation~\eqref{area xi} can be rewritten in terms of expansion $\theta=\nabla_{\mu}u^{\mu}$. Using the definitions of $u^{\mu}$~\eqref{veldef}, $a^{\mu}$~\eqref{acceldef} and $n^{\mu}=\alpha a^{\mu}$, we can modify equation~\eqref{area xi} to read
\begin{equation}
\delta A=\int_{\Sigma}\left(\theta-\frac{3}{2}n^{\mu}n^{\nu}\nabla_{\mu}u_{\nu}\right)\text{d}t\text{d}A.
\end{equation}
If the second term in the integrand vanishes, the formula reduces to
\begin{equation}
\label{area theta}
\delta A=\int_{\Sigma}\theta \text{d}t\text{d}A,
\end{equation}
and both approaches to computing the area yield exactly the same expression. In flat spacetime, it is easy to check that the second term is indeed equal to zero. In a curved spacetime, its value depends on the curvature-dependent ambiguities in the expression~\eqref{u^i_1c} for $u^{i}_{1c}$. Setting $\tilde{C}_3=0$, $\tilde{C}_4=1/6$, leads to $n^{\mu}n^{\nu}\nabla_{\mu}u_{\nu}=0$ as required. Fixing these two constants in no way constraints the two ambiguities in the SLC area we discuss in section~\ref{section: therm}, since these ambiguities parametrized by numbers $C_1$ and $C_2$ are each a combination of four arbitrary parameters (see equations~\eqref{C1} and~\eqref{C2}). Then, fixing two of them imposes no restrictions on the values of $C_1$ and $C_2$.

In conclusion, both expressions for the area of the SLC, equations~\eqref{area xi} and~\eqref{area theta} are operationally equivalent and include the same ambiguities. Then, the results we discuss in section~\ref{section: therm} also directly apply to the derivation of the Einstein equations in the seminal paper~\cite{Parikh_2018}.

%------------------------------------------------
%\renewcommand{\refname}{ References}
\bibliography{article}
%------------------------------------------------

\end{document}